\documentclass[aps,prb,twocolumn,superscriptaddress,showpacs]{revtex4}

\usepackage{amsmath}
\usepackage{bm}
\usepackage{graphicx}

\begin{document}

\title{Quantum electron splitter based on two quantum dots attached to leads}

\author{A.\ V.\ Malyshev}
\thanks{On leave from Ioffe Institute, St. Petersburg, Russia}
\affiliation{Departamento de F\'{\i}sica de Materiales,
Universidad Complutense, E-28040 Madrid, Spain}

\author{P. A. Orellana}
\affiliation{Departamento de F\'{\i}sica, Universidad Cat\'{o}lica
del Norte, Casilla 1280, Antofagasta, Chile}

\author{F.\ Dom\'{\i}nguez-Adame}
\affiliation{Departamento de F\'{\i}sica de Materiales,
Universidad Complutense, E-28040 Madrid, Spain}

\begin{abstract}

Electronic transport properties of two quantum dots side-coupled
to a quantum wire are studied by means of the two impurity
Anderson Hamiltonian. The conductance is found to be a
superposition of a Fano and a Breit-Wigner resonances as a
function of the Fermi energy, when the gate voltages of the
quantum dots are slightly different. Under this condition, we
analyze the time evolution of a Gaussian-shaped superposition of
plane waves incoming from the source lead, and found that the wave
packet can be splitted into three  packets at the drain lead. This
spatial pattern manifests in a direct way the peculiarities of the
conductance in energy space. We conclude that the device acts as a
quantum electron splitter.

\end{abstract}


\maketitle

\section{Introduction}

Quantum dots (QDs) are often referred to as \emph{artificial
atoms\/} since they present discreteness of energy and charge.
Moreover, progress in nanofabrication of quantum devices enabled
to form an artificial molecule sharing electrons from two or more
QDs. In view of the analogy of QDs-based systems and atomic
systems, new ways to look for electronic effects are being
explored. In this regard, it has been recently demonstrated that
coupled QDs shows the electronic counterpart of Fano and Dicke
effects that can be controlled via a magnetic
flux.~\cite{Orellana04} In the case of QDs, Fano resonance
coexists with Coulomb interaction, giving rise to new quantum
transport regimes.~\cite{Johnson04} In addition, Fano resonances
have been clearly observed in a quantum wire (QW) with a
side-coupled QD, and it has been proven that this geometry can be
used as an accurate interferometer.~\cite{Kobayashi04,Sato05}

More recently, we have considered electron transport in a double
QD side attached to a QW  by means of the two impurity Anderson
Hamiltonian.~\cite{Orellana06} The conductance was found to be a
superposition of a Fano and a Breit-Wigner resonances as a
function of the Fermi energy, provided the gate voltages of the
QDs were slightly different. Remarkably, previous numerical
simulations (not shown here) demonstrated that the
electron-electron interaction  is not responsible of this
phenomenon, which can be then regarded as a one-electron process.

In this work we provide further progress along this direction. To
be specific, we solve analytically the scattering by the QDs of a
wide electronic wave packet moving in the QW, and find that the
wave packet can be splitted into three packets after the
scattering process, one delayed with respect to the other.
Remarkably,  this spatial pattern manifests in a direct way the
peculiarities on the conductance in energy space. The analytical
approach to the problem allows us to understand the dependence of
the scattering event on the various parameters of the model. As a
result, a fine control of the electron dynamics in the QW can be
achieved by changing the gate voltages of the two attached QDs.

\section{Conductance at zero temperature} \label{conductance}

\begin{figure}[ht]
\centerline{\includegraphics[width=45mm,angle=0,clip]{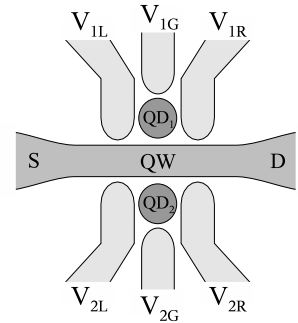}}
\caption{Schematic view of the two quantum dots attached to
quantum wire. Current passing from the source (S) to the drain (D)
is controled by the gate voltages $V_{1G}$ and $V_{2G}$.}
\label{fig1}
\end{figure}

The system under consideration is formed by two QDs connected to a QW waveguide,
as shown schematically in Fig.~\ref{fig1}. The full system is modeled by a two
impurity Anderson Hamiltonian, that can be written as~\cite{Orellana06}
\begin{eqnarray}
{\mathcal H}&=&-v\sum_{\langle i\neq j\rangle}\,(c_{i}^{\dagger}
c_{j}^{}+c_{i}^{}c_{j}^{\dagger}) \nonumber \\
&-&V_{0}\sum_{\alpha}(d_{\alpha}^{\dagger}c_{0}^{}+c_{0}^{\dagger}d_{\alpha}^{})
+\sum_{\alpha}\varepsilon_{\alpha}d_{\alpha}^{\dagger} d_{\alpha}^{}\ ,
\end{eqnarray}
where $c_{i}^{\dagger}$  is the creation operator for an electron
at site $i$ of the QW, $d_{\alpha}^{\dagger}$ is the corresponding
operator for an electron in the QD $\alpha=1,2$. The on-site
energy of the QW is assumed to be zero and the hoppings are taken
to be $v$. The hopping $V_{0}$ couples the QDs to the QW. Notice
that spin indices are omitted hereafter since we do not consider
applied magnetic fields or any other interaction breaking
spin-degeneracy.

The linear conductance can be obtained from the Landauer formula at zero
temperature
\begin{equation}
\mathcal{G}=\frac{2e^2}{h}\,T(\omega=\varepsilon_F)\ ,
\end{equation}
where $T(\omega)$ is the transmission probability ($T=|t|^2$) with
$t$ the transmission amplitude. The transmission amplitude is
given by \cite{Fisher81}
\begin{equation}
t(\omega)=2iv\sqrt{1-\omega^2/4v^2} \,G^W_{0}\ .
\label{transmission}
\end{equation}
Here $G^W_{0}$ is the Green's function at site $0$ of the QW.

By using the Dyson equation we calculate the Green's function at
site $0$ of the QW coupled to the QDs, obtaining the following
expression:
\begin{equation}
G_{0}^{W}=\frac{1}{2iv\sin k}\ \frac{1}{1-i\eta (g_1+g_2)}\ ,
\label{afterDyson}
\end{equation}
where $\eta= V_0^2/2v$,  $k=\arccos(-\omega/2v)$, and
$g_{\alpha}=1/(\omega - \epsilon_{\alpha})$ ($\alpha =1,2$). The
transmission amplitude can be obtained from
Eq.~(\ref{transmission}),
\begin{equation}
t(\omega)=\frac{(\omega-\varepsilon_1)(\omega-\varepsilon_2)}%
{(\omega-\varepsilon_1)(\omega-\varepsilon_2)-i\eta
(2\omega-\varepsilon_1-\varepsilon_2)}\ .
\label{afterDyson2}
\end{equation}
Finally, we arrive at the following expression for the linear
conductance at zero temperature
\begin{equation}
\mathcal{G}=
\frac{(2e^2/h)(\varepsilon_F-\varepsilon_1)^2(\varepsilon_F-\varepsilon_2)^2}
{(\varepsilon_F-\varepsilon_1)^2(\varepsilon_F-\varepsilon_2)^2
+\eta^2(2\varepsilon_F-\varepsilon_1-\varepsilon_2)^2}\ .
\end{equation}

The density of states of the QDs can give us a better
understanding of the transport properties of the system. To obtain
it we calculate the diagonal elements of the Green's functions of
the QDs. In doing so we obtain\cite{Orellana06}
\begin{equation}
\rho=\sum_{\alpha}
\frac{(\eta/\pi)(\omega-\varepsilon_{\alpha})^2}{(\omega-\varepsilon_1)^2
(\omega-\varepsilon_2)^2+\eta^2
(2\omega-\varepsilon_1-\varepsilon_2)^2}\ .
\end{equation}
Setting the sites energies as $\varepsilon_1=\varepsilon_0+\Delta
V$ and $\varepsilon_2=\varepsilon_0-\Delta V$ and taking $\Delta V
\ll \eta$, we get
\begin{equation}
\rho\approx\frac{2\eta/\pi}{(\omega-\varepsilon_0)^2+4\eta^2} +
\frac{\Delta V^2/2\eta\pi}{(\omega-\varepsilon_0)^2+(\Delta
V^2/2\eta)^2}\ . \label{dos}
\end{equation}
The density of states is the sum of two Lorentzians with widths
$\Gamma_+ = 2\eta$ and $\Gamma_{-}=\Delta V^2/2\eta$. On the other
hand the conductance can be written as
\begin{equation}
\mathcal{G}=\frac{(2e^2/h)\varepsilon_F^2}
{(\varepsilon_F-\varepsilon_0)^2+4\eta^2}+
\frac{(2e^2/h)(\Delta V^2/2\eta)^2}
{(\varepsilon_F-\varepsilon_0)^2+(\Delta V^2/2\eta)^2} \ .
\end{equation}
The conductance is the superposition of the a Fano line shape and
a Breit-Wigner line shape. In the limit $\Delta V \rightarrow 0$
one bound state appears
\begin{eqnarray}
\rho&=&\frac{1}{\pi}\,\frac{2\eta}{(\omega-\varepsilon_0)^2+4\eta^2}+
\delta (\omega-\varepsilon_0)\ ,
\label{rho}
\end{eqnarray}
since one of the states is decoupled from the continuum. A similar
effect is already observed in two-channel resonance
tunneling.~\cite{Shahbazyan} In this limit, the conductance is
reduced to a Fano line shape
\begin{equation}
\mathcal{G}=\frac{2e^2}{h}\frac{\varepsilon_F^2}
{(\varepsilon_F-\varepsilon_0)^2+4\eta^2}\ .
\end{equation}

The bound state arises from the indirect coupling of both QDs
through the QW, giving rise to level mixing and formation of
collective anti-bonded states which have zero amplitude at the QW
and are therefore decoupled from it. This mixing arises from
quantum interference in the transmission through the two different
discrete states (the QD levels) coupled to leads. This result is
similar to the Dicke effect in optics, taking place in the
spontaneous emission of two closely-lying atoms radiating a photon
into the same environment.\cite{dicke} The coherent indirect
coupling of the two energy levels gives rise to the splitting of
the decay rates (level broadening), into a fast (superradiant) and
a slow (subradiant) mode.\cite{brandes} Under more realistic
experimental conditions when the QW wire has a finite width such
state would couple weakly to the QW giving rise to broadening of
the $\delta$-function in~(\ref{rho}).

\section{Time-dependent electron dynamics}

Having discussed the steady state properties of the system, we now consider the
time-dependent dynamics of electron wave packets. The initial wave packet was
chosen to be a narrow Gaussian of width $a$ in real space, moving in the QW
\begin{eqnarray}
\psi_0(x,t) &=&
\frac{\psi_0}{\sqrt{1 + i\hbar t/m a^2}}
\exp{ \left[ - \frac{(x-v_0t)^2}{2  a^2(1+i\hbar t/m a^2) }\right]}
\nonumber\\
& \times &
\exp{\left[i k_0 x - i \frac{\hbar k_0^2 }{2m}\, t \right]} \ .
\label{initial}
\end{eqnarray}
Here the group velocity is $v_0=\hbar k_0/m$ and the kinetic energy of
the pulse is $\varepsilon_P = \hbar^2 k_0^2 /2m$.

For the sake of clarity, we introduce the following dimensionless variables
$\xi=x/a$ and $\tau=\hbar t/ma^2$, as well as the parameter $\omega_0=k_0a$.
In doing so, the wave packet can be rewritten as follows
\begin{eqnarray}
\psi_0(\xi,\tau) &=& \frac{\psi_0}{\sqrt{1 + i\tau}}
\exp{ \left[ - \frac{(\xi -\omega_0\tau)^2}{2(1+i\tau) }\right]}
\nonumber\\
& \times &
\exp{\left[i \omega_0 \xi - i \omega_0^2\tau/2\right]} \ .
\label{initial_reduced}
\end{eqnarray}

Following the approach introduced in Ref.~\onlinecite{Wulf05}, the transmitted
wave packet after the scattering by the QDs is obtained as
\begin{subequations}
\begin{equation}
\psi(\xi > 0,\tau) \sim \int_{0}^{\infty} dq\, \widetilde{\psi}(q)\, t(q)\,
e^{\left[i q \xi - i q^2\tau/2\right]} \ .
\label{transmitted}
\end{equation}
From Eq.~(\ref{afterDyson2}) we get approximately
\begin{equation}
t(q)=1+\frac{i\Gamma}{q-\omega_0-i\Gamma}-\frac{i\gamma}{q-\omega_0+i\gamma}\ ,
\label{afterDyson2_bis}
\end{equation}
\end{subequations}
where we set $\varepsilon_P=\varepsilon_0$. Here
$\Gamma=2\eta/\varepsilon_0\omega_0$ and
$\gamma=\Delta V^2/2\eta\varepsilon_0\omega_0$. After performing the integration
in~(\ref{transmitted}), the transmitted wave packet is found to be
\begin{subequations}
\label{three_terms}
\begin{equation}
\psi(\xi > 0,\tau) \sim \psi_0(\xi,\tau)+\psi_1(\xi,\tau)+\psi_2(\xi,\tau) \ ,
\label{transmitted2}
\end{equation}
where $\psi_0(\xi,\tau)$ is the initial wave packet~(\ref{initial}) and
\begin{eqnarray}
\psi_1(\xi,\tau) &=& \sqrt{\frac{\pi}{2}}\,\gamma \,
\exp{\left[i \omega_0 \xi - i \omega_0^2\tau/2\right]} \nonumber \\
&\times & \exp{\left[\gamma (\xi -\omega_0\tau) + \frac{\gamma^2}{2}\,
(1+i\tau)\right]}\nonumber \\
&\times & \mathrm{erfc}
   \left[
      \frac{\xi -\omega_0\tau+\gamma(1+i\tau)}{\sqrt{2}\sqrt{1+i\tau}}
   \right] \ ,
\end{eqnarray}
\end{subequations}
erfc being the complementary error function. Moreover, $-\psi_2(\xi,\tau)$ is
obtained from $\psi_1(\xi,\tau)$ simply replacing $\gamma$ by $\Gamma$.
Therefore, it becomes apparent that the scattered wave packet originates from
the interference of three contributions, resulting in a complicated spatial
pattern as we will show below.

\begin{figure}[ht]
\centerline{\includegraphics[width=70mm,angle=0,clip]{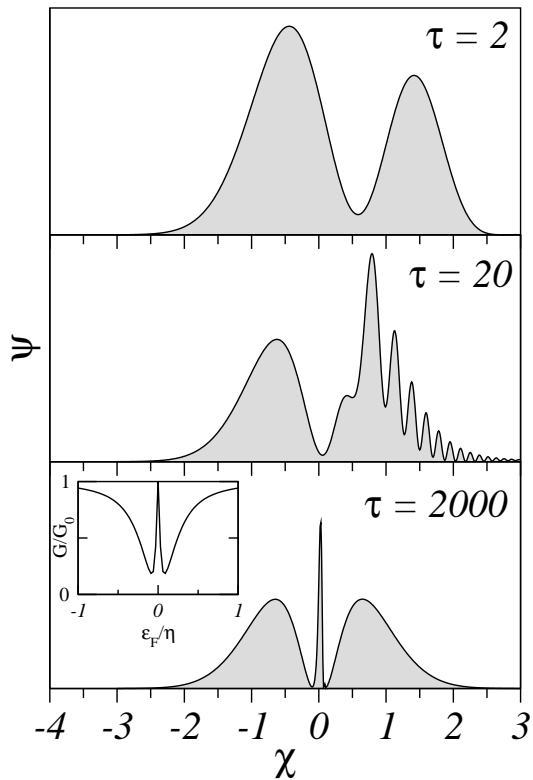}}
\caption{Scattered wave packet at three different values of the
dimensionless time $\tau$, indicated on each panel. Vertical axis
is scaled by a factor $\sqrt{1+\tau^2}$ and
$\chi=(\xi-\omega_0\tau)/\sqrt{1+\tau^2}$. The inset shows the
corresponding conductance, in units of $G_0=2e^2/h$, as function
of the Fermi energy, in units of the effective coupling $\eta$.}
\label{fig2}
\end{figure}

Figure~\ref{fig2} depicts the scattered wave packet at three different values of
the dimensionless time $\tau$, indicated on each panel. The chosen parameters
are $\gamma=0.02$ and $\Gamma=0.5$. Hereafter we set $\epsilon_0=2v$, that is we
consider a wave packed centered at the bottom of the QW band where the energy
dispersion is quadratic. To improve clarity, both vertical and horizontal axis
were scaled, as dispersion makes the wave packet lower and wider on increasing
time. We observe that at earlier stages of the scattering process the wave
packet splits into two pulses, but after interference with the reflected one,
finally a third and much narrowed peak appears at the center.

The width of the third peak is given by $\gamma$, as it can be
seen  from Eq.~(\ref{three_terms}). To show this, one can use the
asymptotic form of the complementary error
function\cite{Abramowitz64} $ \mathrm{erfc}(z)\approx
e^{-z^2}/z\sqrt{\pi}$ ($z\gg 1$) for $\xi \sim \omega_0 \tau$ and
$\gamma \sqrt{\tau},\, \Gamma\sqrt{\tau}\gg 1$ to obtain
\begin{eqnarray}
\psi(\xi > 0,\tau) & \sim & \psi_0(\xi,\tau)
   \left[
      1+\frac{\gamma(1+i\tau)}{\xi-\omega_0\tau+\gamma(1+i\tau)}
   \right.
   \nonumber \\
      &-& \left.\frac{\Gamma(1+i\tau)}{\xi-\omega_0\tau+\Gamma(1+i\tau)}
   \right] \ .
\label{asymptotic}
\end{eqnarray}
The final pattern can be explained from this asymptotic result.
The scattered wave packet has two additive contributions, the
initial one and a very narrow central peak [two first terms
of~(\ref{asymptotic})], while the third term is negative, causing
the central and wide dip.

\begin{figure}[ht]
\centerline{\includegraphics[width=70mm,angle=0,clip]{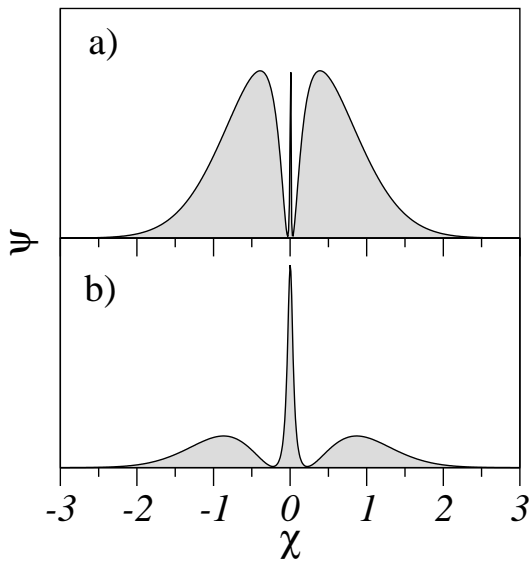}}
\caption{Scattered wave packet at dimensionless time $\tau=20000$ for a)
$\gamma=0.0045$ and $\Gamma=0.11$, and b)~$\gamma=0.045$ and $\Gamma=1.1$.
Vertical axis is scaled by a factor $\sqrt{1+\tau^2}$ and
$\chi=(\xi-\omega_0\tau)/\sqrt{1+\tau^2}$.}
\label{fig3}
\end{figure}

Thus we came to the conclusion that the appearance of the transmitted pulse can
be controlled by varying $\eta$, namely the effective coupling of the two QDs
and the QW,  as well as their voltage ($\Delta V$). This result allows for a
fine control of the transport properties of the QW with adjusting the parameters
of the two QDs. As an example, Fig.~\ref{fig3} the asymptotic wave packet
(dimensionless time $\tau=20000$) for two sets of parameters: a)~$\gamma=0.0045$
and $\Gamma=0.11$, and b)~$\gamma=0.045$ and $\Gamma=1.1$. One can notice a
change in the spatial pattern by varying the widths of the central peak and the
central dip.

\begin{figure}[ht]
\centerline{\includegraphics[width=70mm,angle=0,clip]{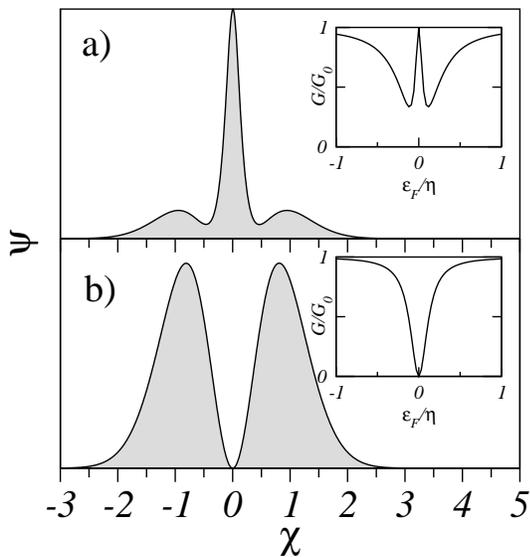}}
\caption{Scattered wave packet at dimensionless time $\tau=2000$
for a) two QDs and b) one QD attached to the QW. Vertical axis is
scaled by a factor $\sqrt{1+\tau^2}$ and
$\chi=(\xi-\omega_0\tau)/\sqrt{1+\tau^2}$. The insets show the
corresponding conductance, in units of $G_0=2e^2/h$, as function
of $\varepsilon_F/\eta$.} \label{fig4}
\end{figure}

The results discussed above refer to two QDs attached to a QW. It is interesting
to elucidate the main differences appearing in the scattered pulse with respect
to the case of a single QD attached to a QW. Figure~\ref{fig4} shows the
asymptotic wave packet (dimensionless time $\tau=2000$) in both cases. It is
clearly seen the vanishing of the central peak after removing one of the two
QDs. Notice that this behavior directly reflects the peculiarities of the DOS
discussed in Sec.~\ref{conductance}, in the sense that the narrow Lorentzian
peak [second term in Eq.~(\ref{dos})] vanishes when one QD is removed.

\section{Conclusions}

We have studied coherent transport through a QW side attached to two QDs with
slightly different gate voltages. To this end, a noninteracting two impurity
Anderson Hamiltonian was used to describe the electronic properties of the
system. We found closed expressions or both the conductance  at zero temperature
and the density of states. The conductance is the superposition of a Fano line
shape and a Breit-Wigner line shape. Also, we found that the density of states
is the sum of two Lorentzians of different widths. It resembles the  Dicke effect
observed in quantum optics. More important, their widths can be controlled via
the coupling of the QDs and the QW.

In addition, we analytically solved the scattering of a Gaussian
wave packet impinging over the QDs. The scattered wave packet was
shown to be the superposition of three pulses, giving rise a
complicated pattern as a  function of time. In spite of this
complex behavior, the long-time  asymptotic pulse reflects the
peculiarities found in the density of states, showing a central
peak whose width can be controlled via the gate voltages. This
peak is absent when one of the QD is disconnected from the QW, and
the time-dependent current after pulse excitation presents two
bumps delayed an amount of $ma^2/\hbar$. For couplings between the
QDs and the QW of the order of few tenths of meV, and $a \sim
20\,$nm, the delay time might be of the order of few ps in GaAs.
To conclude, this phenomenon opens possibilities to design new
quantum electron devices like electron splitters, based on
physical effects that are usually encountered in quantum optics.

\acknowledgments

Work at Madrid was supported by MEC (MAT2003-01533). A.\ V.\ M.
acknowledges financial support of MEC through the Ram\'{o}n y
Cajal programm. P.\ A.\ O.\ would like to thank financial support
from Milenio ICM P02-054-F and FONDECYT under Grants 1060952 and
7020269. Moreover, P.\ A.\ O.\ would like to thank the hospitality
of the Departamento de F\'{\i}sica de Materiales of the
Universidad Complutense de Madrid during his visit.

\end{document}